\documentclass[10pt,twocolumn,letterpaper]{article}

\usepackage{ijcb}
\usepackage{times}
\usepackage{epsfig}
\usepackage{graphicx}
\usepackage{amsmath}
\usepackage{amssymb}
\usepackage[norule,symbol,perpage]{footmisc}

\usepackage{subfigure}
\usepackage{url}

\ijcbfinalcopy 

\ifijcbfinal\fi

\makeatletter
\def\ps@IEEEtitlepagestyle{
\def\@oddfoot{\mycopyrightnotice}
\def\@evenfoot{}
}
\def\mycopyrightnotice{
{\hfill \footnotesize 978-1-7281-9186-7/20/\$31.00 \copyright 2020 IEEE\hfill}
}
\makeatother

\begin{document}

\title{Recognition Oriented Iris Image Quality Assessment in the Feature Space}

\author{
Leyuan Wang, Kunbo Zhang, Min Ren, Yunlong Wang, Zhenan Sun\\
School of Artificial Intelligence, UCAS\\
Center for Research on Intelligent Perception and Computing\\
National Laboratory of Pattern Recognition, CASIA\\
{\tt\small \{leyuan.wang, min.ren, yunlong.wang\}@cripac.ia.ac.cn}, {\tt\small kunbo.zhang@ia.ac.cn}, {\tt\small znsun@nlpr.ia.ac.cn}
}

\maketitle
\thispagestyle{empty}

\begin{abstract}
   A large portion of iris images captured in real world scenarios are poor quality due to the uncontrolled environment and the non-cooperative subject. To ensure that the recognition algorithm is not affected by low-quality images, traditional hand-crafted factors based methods discard most images, which will cause system timeout and disrupt user experience. In this paper, we propose a recognition-oriented quality metric and assessment method for iris image to deal with the problem. The method regards the iris image embeddings Distance in Feature Space (DFS) as the quality metric and the prediction is based on deep neural networks with the attention mechanism. The quality metric proposed in this paper can significantly improve the performance of the recognition algorithm while reducing the number of images discarded for recognition, which is advantageous over hand-crafted factors based iris quality assessment methods. The relationship between Image Rejection Rate (IRR) and Equal Error Rate (EER) is proposed to evaluate the performance of the quality assessment algorithm under the same image quality distribution and the same recognition algorithm. Compared with hand-crafted factors based methods, the proposed method is a trial to bridge the gap between the image quality assessment and biometric recognition.
\end{abstract}

\let\thefootnote\relax\footnotetext{\mycopyrightnotice}

\section{Introduction}

\begin{figure}[h]
   \begin{center}
      \subfigure[Hand-crafted factors based framework]{
         \includegraphics[width=0.9\linewidth]{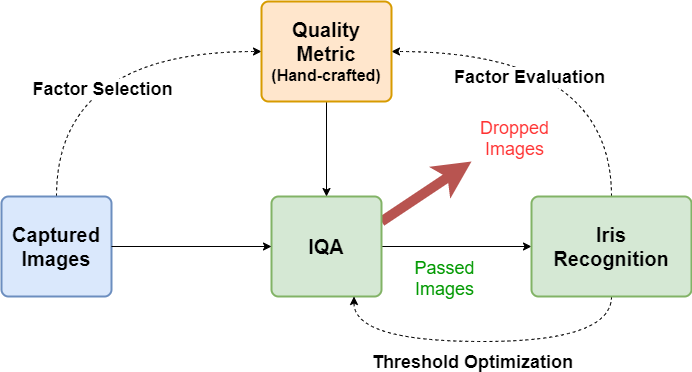}
         \label{fig:hdframework}
      }

      \subfigure[DFS based framework]{
         \includegraphics[width=0.9\linewidth]{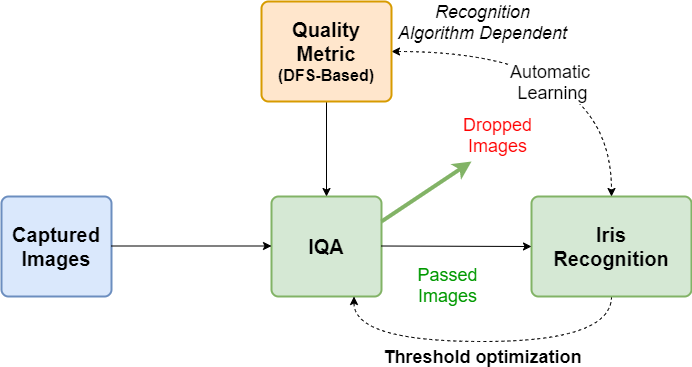}
         \label{fig:dfsframework}
      }
   \end{center}
   \caption{Iris quality assessment framework}
   \label{fig:framework}
\end{figure}

Iris recognition has been found as one of the most accurate and reliable biometric approaches for personal identification applications. Existing state-of-the art iris recognition algorithms are very successful in feature extraction as long as there is an high quality image. However, low quality iris image has shown to reduce the performance of iris location, segmentation and recognition. Therefore, accurate and objective assessment of iris image quality will improve iris recognition performance by dropped low quality image, especially for non-ideal situations such as non-cooperative, long-distance, multi-person etc.

Previous work on iris image quality assessment can be categorized as multi-step method and fusion method. The multi-step method is the assessment of different quality metrics across iris preprocessing to discard the image with score below the designed thresholds step by step. Fusion method base on certain rules and functions to fuse multiple quality metrics as a single final score during iris preprocessing. Both multi-step and fusion methods are designed and optimized manually for quality factors. The framework is shown in Figure \ref{fig:hdframework}. There are many quality factors affecting images quality assessment, such as occlusion, blur, off-angle, resolution, lighting variation, etc. \cite{Tabassi_Grother_Salamon_2011}

In general, iris image quality assessment can be regarded as a no-reference image quality assessment (NR-IQA) problem which focuses on realistic scenario. Different from full-reference or reduced-reference image quality assessment methods in which reference image is available, NR-IQA method tends to extract features and quantify distortion from target image while reference image is unavailable. Recently, with the success of convolutional neural network (CNN) in computer vision, researchers explored the potential of applying CNN to NR-IQA \cite{Kang_2014_CVPR,Liu_Weijer_Bagdanov_2017}. A simple and effective method is to use CNN to obtain a local quality assessment and a weighted pooling strategy to estimate the global quality score \cite{Gu_Meng_Xiang_Pan_2019}. The weights are generated by saliency prediction or generic object detection methods.

Existing iris quality assessment methods, both of multi-step or fusion, are based on hand-crafted quality factors. these quality metrics are designed according to human experience and their scores are consistent with human's subjective perception. Although it is intended to assist iris recognition, the relationship between such factors and recognition performance \cite{kalka2010estimating} is not thoroughly considered during metrics design. Hence, it is questionable that these hand-designed factors are optimal measures for iris quality assessment. At the same time, the calculation of some factors such as motion blur is very complicated \cite{kang2008restoration,li2011comprehensive, proencca2010quality}, and
time consuming. Unlike the natural image quality assessment, an ideal is quality metrics of iris image should be able to predict the performance of the recognition algorithm and independent of human subjective experience. Therefore, an efficient recognition-oriented iris quality assessment method is worth to research.

In this paper, we propose an iris image quality metrics and an iris image assessment framework towards better recognizing iris images for personal identification as shown in Figure \ref{fig:dfsframework}. Given a specific iris recognition algorithm, we claim that there is is only one embedding in the feature space for any human iris theoretically. The feature space distance between the embeddings of the verification image and the enrollment image can be used as the iris image quality measurement. We present an attention-based pooling strategy that uses the coarse iris segmentation results as weights and predicts feature space distance.

The main contributions of this paper are as follows:

1. We explore and establish the relationship between iris quality assessment and iris recognition performance. The embeddings distance in feature space (DFS) is introduced as the quality metric, which is a parameter-free architecture compared with traditional methods using hand-crafted parameters.

2. We introduce the definition of image rejection rate (IRR) which is related to the possibility of iris recognition system timeout. The proposed IRR-EER curve is able to quantitatively evaluate the performance of iris image quality assessment algorithms.

3. An attention-based pooling strategy is presented, which predicts the quality score of the iris image using deep neural networks to replace the traditional hand-crafted methods. This network architecture has been verified on our non-ideal iris database with a promising result.
\section{Related Work}

\begin{figure}[h]
   \begin{center}
      \subfigure[Uncontrolled environment]{
         \includegraphics[width=0.3\linewidth]{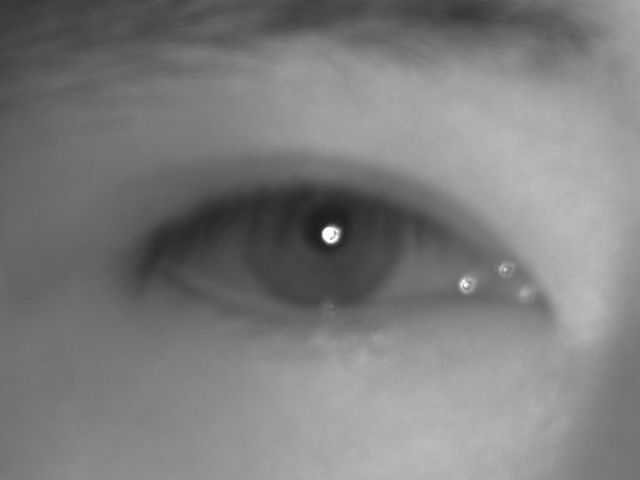}
         \includegraphics[width=0.3\linewidth]{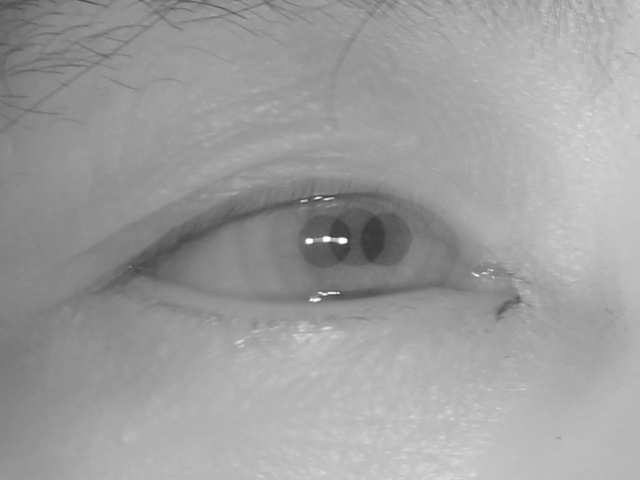}
         \includegraphics[width=0.3\linewidth]{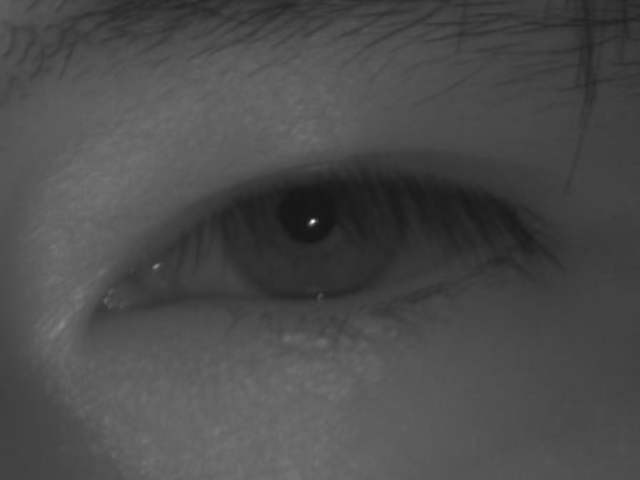}
         \label{fig:lqiris_a}
      }

      \subfigure[Non-cooperative subject]{
         \includegraphics[width=0.3\linewidth]{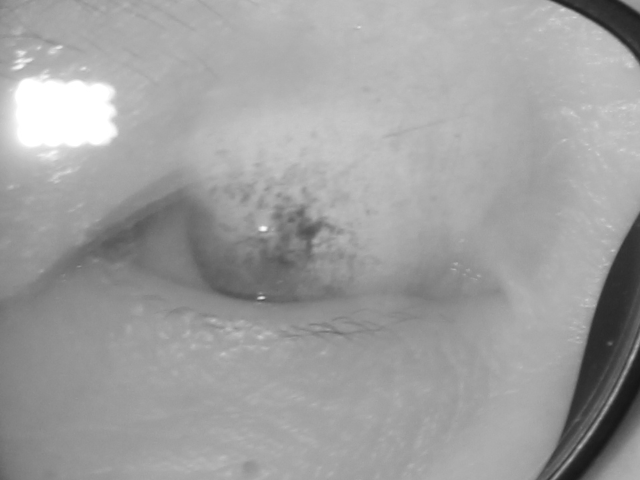}
         \includegraphics[width=0.3\linewidth]{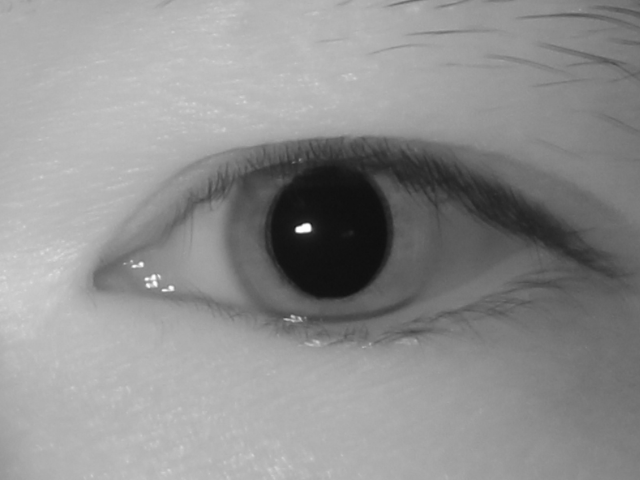}
         \includegraphics[width=0.3\linewidth]{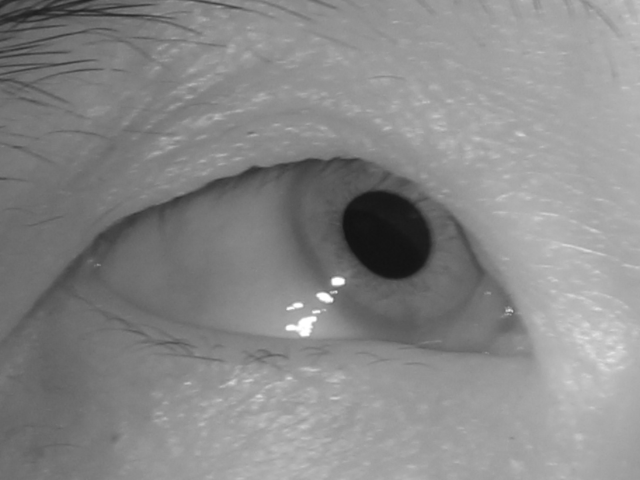}
         \label{fig:lqiris_b}
      }
   \end{center}
   \caption{Low quality iris images}
   \label{fig:lqiris}
\end{figure}

In biometric recognition, whether the quality of biometric samples can be accurately evaluated is an important issue, which affects the performance of subsequent recognition process. Compared with face and fingerprint, the quality assessment method of iris is more complicated. On one hand, iris image acquisition is affected by environmental factors and acquisition equipment. The originally collected iris image is prone to distortions at image level, such as image resolution, defocus blur, motion blur, and lighting (Figure \ref{fig:lqiris_a}). On the other hand, the iris itself will also undergo disturbances due to subject variations such as occlusion, pupil dilation, and off-angle (Figure \ref{fig:lqiris_b}).

\begin{table*}[h]
   \begin{center}
      \begin{tabular}{|p{0.2\textwidth}|p{0.35\textwidth}|p{0.35\textwidth}|}
         \hline
         \textbf{Quality factor} & \textbf{Impairment}                         & \textbf{Evaluation algorithm}                                                       \\
         \hline
         Sharpness               & Defocus, Compression                        & Based on gradient changes \cite{zhang1999method}                                    \\
                                 &                                             & Based on global spatial filtering \cite{daugman2009iris,wan2007iris,wei2006robust}  \\
                                 &                                             & Based on region of interest filtering \cite{kalka2010estimating,pan2007algorithm}   \\
         \hline
         Motion blur             & Distortion due to motion                    & Based on spectrum direction analysis \cite{kalka2010estimating,li2011comprehensive} \\
         \hline
         Off-angle               & Optical axis of camera and eye not lined up & Based on pupil circle model \cite{kalka2010estimating}                              \\
                                 &                                             & Based on corneal reflex position \cite{li2011comprehensive}                         \\
         \hline
         Usable iris             & Occlusion (reflections, eye-wear. etc.)     & Based on rough location \cite{kalka2010estimating}                                  \\
                                 &                                             & Based on fine location \cite{li2011comprehensive}                                   \\
         \hline
         Gray level spread       & Illumination, Saturation                    & Based on local gray differences \cite{kalka2010estimating,zuo2009global}            \\
                                 &                                             & Based on information entropy \cite{belcher2008selective}                            \\
         \hline
         Dilation                & Ambient light, Intrinsic                    & Based on the ratio of the inner and outer radius of the iris \cite{zuo2009global}   \\
         \hline
      \end{tabular}
   \end{center}
   \caption{Hand-crafted quality metrics and assessment methods}
   \label{tab:hand-crafted method}
\end{table*}

The commonly used iris quality assessment methods can be placed into two categories, one is multi-step method, another is fusion method. Multi-step method focuses on several quality metrics. If the score of any quality metric is below the threshold, the iris image is discarded. Otherwise, the next quality metric is calculated until all quality metrics have been checked \cite{Kalka_2005,Peng_Wu_2015,wei2006robust}. This kind of method can effectively remove low-quality iris images with prominent single-quality distortion, and is closely integrated with the iris recognition process. However, it is not sensitivity to distortion in iris image quality under the combined effects of multiple measures. Common quality metric and assessment methods are shown in Table \ref{tab:hand-crafted method}. Fusion method combines quality metrics according to certain rules and outputs an overall quality score. Craig \cite{belcher2008selective} and Aditya \cite{abhyankar2009iris} used function mapping to fuse individual quality measure scores to obtain the final quality score. Nathan \cite{kalka2010estimating} used Dempstefr-Shafer fusion theory to fuse multiple quality metric, the premise of which is to distinguish two subsets of good and bad quality as hypothetical propositions used in training. Mohsen et al. \cite{Jenadeleh_Pedersen_Saupe_2018} proposed an assessment measure based on statistical features of the local difference sign-magnitude transform.

The quality score of the iris image should be able to predict the recognition performance, and the recognition performance for iris images with different quality states should have significant differences. Existing iris quality assessment methods, both of multi-step method or fusion method, are based on hand-crafted quality metrics. Nathan's research \cite{kalka2010estimating} confirms that there is a certain correlation between these quality metrics and the performance of the recognition algorithm, but it does not strictly prove that these factors can directly affect the recognition. And these quality metrics cannot cover all types of quality distortions \cite{Tabassi_Grother_Salamon_2011}. Therefore, the estimation of the iris quality assessment algorithm based on hand-crafted quality metric will have deviation.

In practical applications, the multi-step method needs to set a threshold for each quality metric, which is difficult to implement and has poor generalization. The fusion method is difficult to combine with the iris recognition process \cite{kalka2010estimating,li2011comprehensive}. The quality score can only be calculated after the preprocessing such as segmentation. Segmenting a large number of low-quality images will increase the overall time consumption of iris recognitions \cite{Zhao_Kumar_2015,wang2019joint}. In long-distance non-cooperative scenarios, most of the iris images acquired are of very poor quality \cite{Nguyen_Fookes_Jillela_Sridharan_Ross_2017}. In order to ensure the accuracy of recognition, the iris quality assessment algorithm based on hand-crafted quality metric will set a high threshold and discard a large number of images. This actually leads to an increased likelihood of timeouts or failure to identify. All in all, the existing iris quality assessment algorithms are difficult to complete long-distance, non-cooperative scene iris recognition tasks which is the key issues of iris recognition research.

\section{Methods}

\subsection{Quality metric toward recognition}

\begin{figure*}[h]
   \begin{center}
      \includegraphics[width=0.95\linewidth]{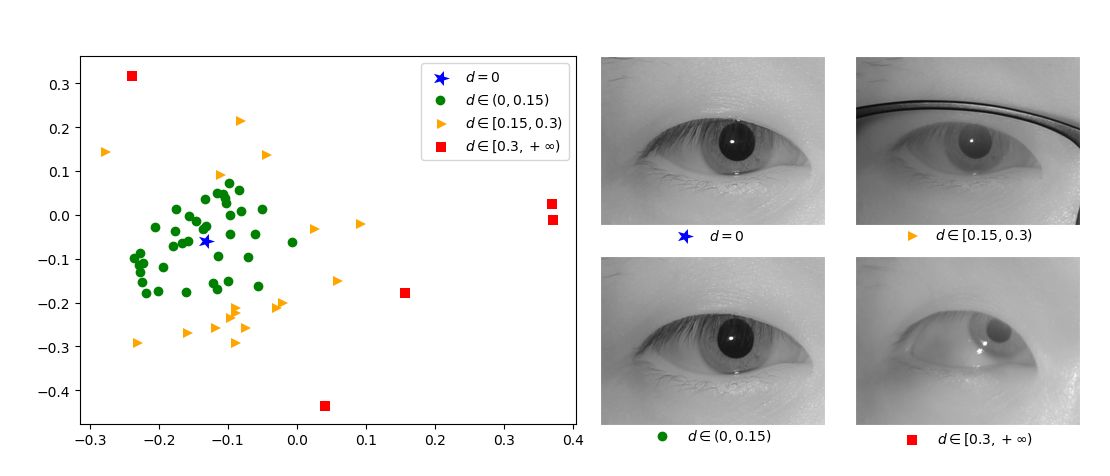}
   \end{center}
   \caption{Iris embedding in feature space}
   \label{fig:embeddings}
\end{figure*}

A key issue in image quality assessment is how to define the quality metric. Unlike the natural image quality that can be judged based on human subjective perception, the quality metric of the biometric samples should be able to predict the performance of the recognition algorithm. \cite{Grother_Tabassi_2007} In practice, iris quality assessment algorithm usually continuously acquires images from the acquisition device and discards low-quality images until the recognition is complete or timeout. There is a trade-off between performance on the one hand and the possibility of timeout on the other hand. Theoretically, with the increase of the quality threshold, the performance of the recognition system will also improve. But the more images that are discarded, the more likely recognition system timeout and fail to complete recognize, because long wait can seriously disrupt user experience. Hence, from the perspective of the whole recognition system, a quality assessment algorithm that can discard as few images as possible on the basis of guaranteed performance is needed. To this end, we propose an iris quality metric based on distance of embeddings in feature space (DFS).

The deep learning-based iris feature extraction methods map the normalized iris image to a two-dimensional feature template \cite{Zhao_Kumar_2019} or one-dimensional feature vectors \cite{Zhang_Li_Sun_Tan_2018}. In this paper, we refer to template and vector as embedding. Distances of two embeddings in the feature space directly correspond to a measure of iris similarity. In training, one possible method is to uses a loss function such as softmax to train the feature extraction network and the classification network together. Another method is to directly use the metric learning method to train the feature extraction network. Such two approaches have the benefit of minimizing the intra-class spacing and maximizing the inter-class spacing \cite{Deng_Guo_Xue_Zafeiriou_2018}.

We propose such a hypothesis that for an ideal eye image $ x_0 $, the corresponding unique embedding $ y_0 = E (x_0) $ in the feature space is calculated by the iris recognition algorithm $ E (x) $. The influence of factors can only capture non-ideal images $ x $, and the corresponding embedding $ y $ will also be relative to $ y_0 $. With $ E (x) $ fixed, the distance in feature space $ d = || y-y_0 || $ can be used to measure the impact of various subjective and objective factors in acquisition process, that is the actual quality of the iris image. Here, we do not consider the specific implementation of iris recognition algorithm including localization, segmentation and feature extraction, and treat it as a black box.

Figure \ref{fig:embeddings} is the embedding of the iris samples in the feature space, where the blue point are the registration sample we provided and the other points are the verification samples. In the feature space, considering that the $x_0$ is not available in reality and the registration sample is collected under a highly controlled situation, we can treat it approximately as the $x_0$. Verification samples $x$ obtained under uncontrolled conditions and $y$ is distributed around $y_0$ . If the classification boundary in the feature space is a hypersphere with $y_0$ as the center and a radius of 0.2. It is almost impossible to misclassify an image with the quality of it similar to green samples, and it is likely to be misclassified if the quality is close to yellow and red samples. The task of iris image quality assessment is to exclude these images that may be misclassified. Therefore, we take DFS as iris quality metric and use a model to learn the mapping between images and DFS directly from dataset.

\subsection{Attention-based pooling strategy}

\begin{figure*}
   \begin{center}
      \includegraphics[width=0.9\linewidth]{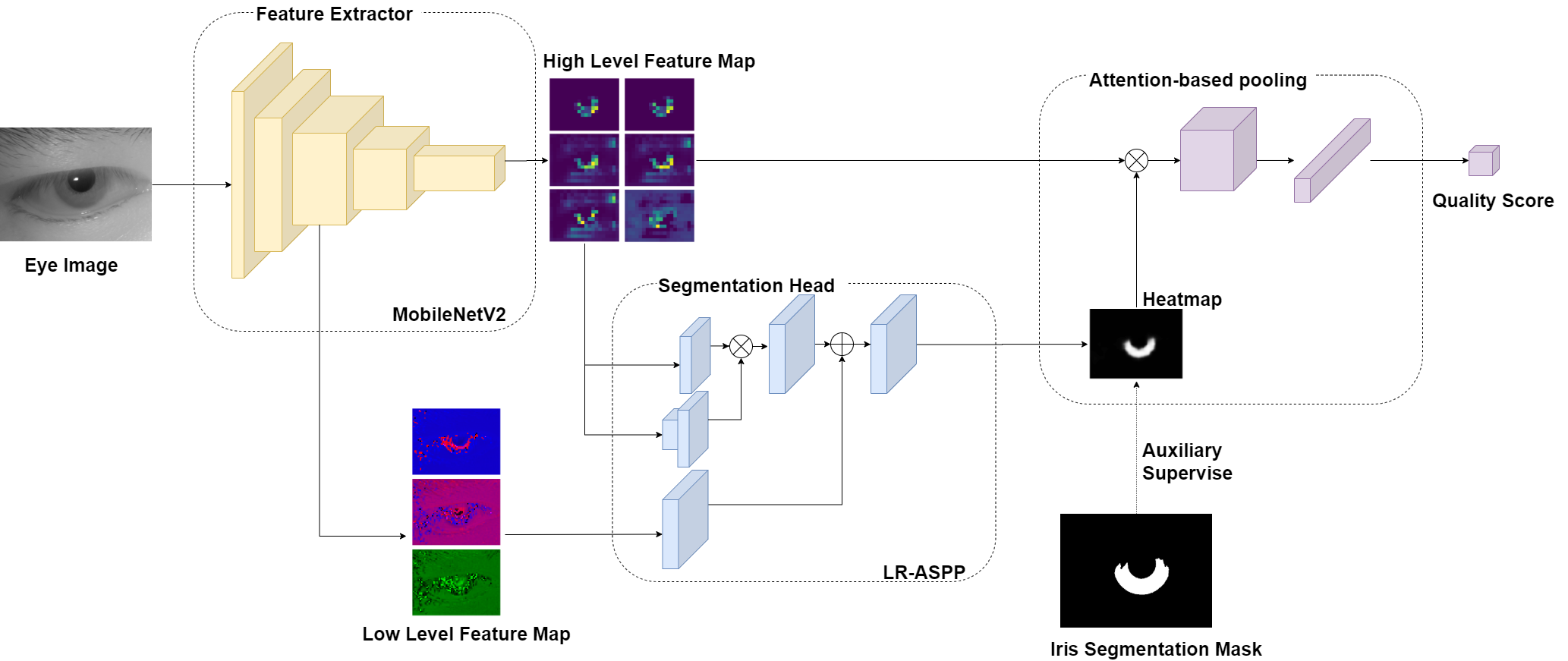}
   \end{center}
   \caption{
      DFS prediction network structure.
      Segmentation module generate a rough segmentation result as the heatmap required by the attention-based pooling module. High-level feature map contains enough information to predict DFS.Multiplying high-level features map with heatmap can suppress the response of non-iris regions.
   }
   \label{fig:netstructure}
\end{figure*}

Natural image quality assessment usually uses a pre-trained classification network as a feature extractor, then adds a global average pooling layer after the last convolution layer to convert the feature map into a quality vector, and finally uses the linear layer to get the global quality score. The structure of the network is shown in Figure \ref{fig:netstructure}. Some researchers \cite{Gu_Meng_Xiang_Pan_2019} use a weighted pooling strategy, whose weight comes from saliency prediction results or target detection results. It is a soft attention mechanism inspired by the human visual system, which makes the image quality assessment algorithm pay more attention to the iris areas of an eye image.

For iris image quality assessment, the quality score calculated by the global average pooling method is greatly affected by the periocular area. For example, the eyelash area contains a lot of high-frequency information, which interferences with the final evaluation result. Inspired by the weighted pooling strategy, we propose a weighted average pooling operation based on the attention mechanism, which multiplies the heatmap with the feature map as weights,

\begin{equation}
   Q(z) = \frac {\sum_{0}^{h}{\sum_{0}^{w}{H(x, y)*F(x, y, z)}}} {\sum_{0}^{w}\sum_{0}^{h}{H(x, y)}}
\end{equation}

\noindent Where quality vector $ Q(z) $ is a c-dimensional vector, heatmap $ H(x, y) $ is (height, width) and feature map $ F(x, y, z) $ is (height, width, channel). The feature map is generated by a quality feature extractor, and the heatmap comes from coarse segmentation of iris which is a specific semantic segmentation problem.

At present, mainstream semantic segmentation networks such as U-net and RefineNet have adopted an encoder-decoder structure. This structure has also succeeded on multiple computer vision tasks. Among them, the encoder gradually reduces the spatial resolution and captures high-level semantic information. The decoder gradually recovers the spatial resolution and reconstruct the segmented image. In practice, feature extraction networks are often used as encoders. So in our task, the quality feature extractor and segmentation encoder use the same classification network. We believe that the high-level semantic information extracted by the semantic segmentation encoder also expresses the quality information of the image, and the reuse of modules can improve the performance and speed of both tasks simultaneously.

\subsection{Criteria of quality assessment algorithm}

In an iris recognition system, low-quality images are discards. The more images that are discarded, the more likely it is to timeout. We introduce the definition of image rejection rate (IRR) which is related to the possibility of iris recognition system timeout. IRR is the ratio of the number of discarded images to the total captured images. The performance of iris recognition algorithm is typically characterized by error rate (EER) which indicates that the false accept rate (FAR) is equal to the false reject rate (FRR). We plotted the relationship between IRR and EER to measure the performance of quality assessment algorithm.

\subsection{Implementation details}

The quality assessment is located at the beginning of the entire iris recognition process. In order to ensure the speed of the entire recognition system, it is necessary to be able to process a large number of images in real time. We propose a light-weight network to achieve this demand. Specifically, the feature extractor tried the pre-trained MobileNetV2 \cite{mbv2} and MobileNetV3 \cite{mbv3}, because it is capable of a large number of visual tasks with few parameters and fast speed running on mobile devices in real-time. Then U-net and LR-ASPP are used as decoder. Although U-net \cite{Ronneberger_Fischer_Brox_2015} was born in 2015, it is still the most widely used model in segmentation tasks due to its simple and effective structure. LR-ASPP was proposed in \cite{mbv3}, which is light-weight and has well performance on Cityscapes dataset. In subsequent experiments, we finally used a combination of MobileNetV2 and LR-ASPP, which have a smaller amount of calculation and fewer parameters while guaranteeing the performance.

We use single eye images which size is $640*480$ as input to generate a normalized prediction quality score ranged from $(0,1)$ and a $160*120$ heatmap. During training, the segmentation result and the prediction result are used as supervision, and the Loss function is set to

\begin{equation}
   Loss = \lambda L_{mask} + (1- \lambda) L_{DFS}
\end{equation}

\noindent Where $ L_{mask} $ is the segmentation loss using cross entropy, and $ L_ {DFS} $ is the prediction loss using MSELoss. Considering that the ultimate purpose of our network is to predict DFS rather than output the coarse segmentation, an annealing method is implemented. Specifically, $ \lambda $ is set to 0.8 at the beginning of training, and then every 50 epoch $ \lambda $ is halved. We train our model with Adam optimizer by setting $ \beta_{1}=0.9 $ and $ \beta_{2}=0.99 $. The learning rate is set to $4*10^{-4}$ initially and is halved four times during training.

\section{Experiments and analysis}

\subsection{Evaluation criteria}

In addition to IRR-EER curve, in our experiments, the performance of iris quality assessment algorithms is evaluated by three widely used criteria, namely linear correlation coefficient (LCC), Spearman rank order correlation coefficient (SROCC) and mean square error (MSE). LCC is the statistical measures of linear relationship between groundtruth and prediction, SROCC measures the monotonicity, and MSE is the standard deviation of the prediction errors.

\subsection{Dataset and recognition algorithm}

Currently, iris datasets such as CASIA-Iris-V4 \cite{CASIA-IrisV4} and ND-IRIS-0405 \cite{Bowyer_Flynn} are collected under controlled containing less distorted images. The UBIRIS \cite{UBIRISv1,UBIRISv2} and WVU \cite{ross2004centralized} datasets contain various low-quality images, but the former was collected under visible light conditions, and the latter has ceased maintenance and is not publicly available.

To further research iris recognition in uncontrolled environment and facing non-cooperative subject, we collected an iris dataset that contains both high and low quality images, named CASIA-Iris-Complex. It consists of two subset, CASIA-Iris-CX1 and CASIA-Iris-CX2. The CASIA-Iris-CX1 is collected at the distance of 0.75m, including pupil dilation, off-angle, occlusion and other distortion factors. The CASIA-Iris-CX2 is collected at distances of 1m, 3m, and 5m, including distortion factors that are prone to occur long-distance and non-cooperative conditions such as blur, exposure, and off-angle.

We take 10,358 images from 164 classes in CASIA-Iris-CX1 and 3,167 images from 58 classes in CASIA-Iris-CX2 as training sets, and the remaining images as two test sets.

\begin{table}[h]
   \begin{center}
      \begin{tabular}{|c|c|c|}
         \hline
         Subset characteristics & CX1    & CX2        \\
         \hline
         Distance               & 0.75m  & 1m, 3m, 5m \\
         No. of images          & 12,980 & 4,249      \\
         No. of classes         & 164    & 74         \\
         No. of Ideal images    & 3,189  & 580        \\
         No. of Nonideal images & 9,791  & 3,714      \\
         \hline
      \end{tabular}
   \end{center}
   \caption{Dataset characteristics}
   \label{tab:dataset}
\end{table}

In order to obtain DFS, our experiments used an state-of-the-art iris segmentation \cite{wang2019joint} and recognition \cite{Zhang_Li_Sun_Tan_2018} algorithms which are pre-trained on CASIA-Iris-V4 and ND-IRIS-0405 datasets. For image that cannot be located and segmented correctly, we manually label and normalize them. On this basis, we extract the embedding of all the images from CASIA-Iris-Complex, and select an image from ideal images as the registered image in each class. And then calculate the cosine similarity between the registered image and all the other images in the same class as DFS.

\begin{table}[h]
   \begin{center}
      \begin{tabular}{|l|l|}
         \hline
         Dataset                       & EER     \\
         \hline
         ND-IRIS-0405                  & 1.39\%  \\
         CASIA-Iris-Thousand           & 1.72\%  \\
         CASIA-Iris-CX1 (all images)   & 12.41\% \\
         CASIA-Iris-CX1 (ideal images) & 1.28\%  \\
         CASIA-Iris-CX2 (all images)   & 28.59\% \\
         CASIA-Iris-CX2 (ideal images) & 16.46\% \\
         \hline
      \end{tabular}
   \end{center}
   \caption{Recognition performance on different datasets}
   \label{tab:dataset}
\end{table}

\subsection{Experimental result}

\begin{figure}
   \begin{center}
      \subfigure[CASIA-Iris-CX1]{
         \includegraphics[width=0.8\linewidth]{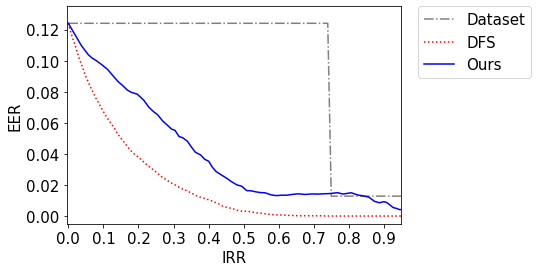}
         \label{fig:eer_irr_SD}
      }
      \subfigure[CASIA-Iris-CX2 subset]{
         \includegraphics[width=0.8\linewidth]{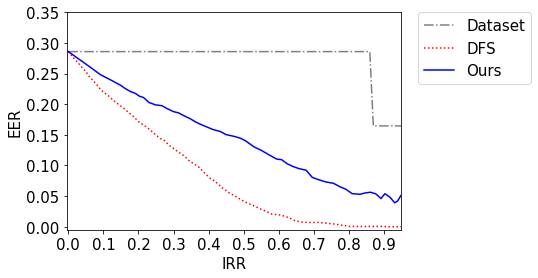}
         \label{fig:eer_irr_TJ}
      }
   \end{center}
   \caption{EER-IRR curve of DFS-based method, (a) is CASIA-Iris-CX1 subset, (b) is CASIA-Iris-CX2.}
   \label{fig:eer_irr}
\end{figure}

First, we plotted the relationship between DFS and EER. As shown in Figure \ref{fig:eer_irr}, the gray dotted line in the figure is a benchmark, which is the result of discarded non-ideal images in the dataset. It can be seen that the recognition performance on the two subsets is improved if low-quality images are discarded. The red solid line is the result of using DFS as quality metric which is the upper bound of quality assessment algorithm . As the IRR increases, the EER gradually decreases, which is consistent with our previous expectations. On the two subsets of close and distant, when the IRR is 0.40 and 0.65, EER reaches 1\%, which is even lower than the result on the ideal image. And the IRR-EER curve of DFS on both subsets is below the curve of benchmark. This shows that during the acquisition process, what we consider to be "non-ideal" still includes a large number of images that can be effectively identified, and the "ideal" image also contains quality metrics that we do not understand and reduce recognition performance. Respectively, the IRR-EER curve reaches an elbow point, when the IRR reached 0.5 and 0.65 respectively while the rate of EER decline is slowed significantly. We believe that the main factor affecting recognition performance before this point is image quality, and the performance improvement after excluding these low-quality images mainly depends on the improvement of algorithms.

After that, the correlation between the model prediction and DFS is also analyzed. The results are shown in Table \ref{tab:correlation_analysis}. Similarly, we also plotted IRR-EER curve of the prediction, as shown in the Figure \ref{fig:eer_irr}. The blue solid line in the figure is our prediction. It can be noticed that prediction curve changes synchronously with DFS curve, and is always below the curve of the benchmark. However, it can also be seen that there is still a certain error between our prediction curve and DFS curve. Therefore, how to improve the accuracy of model is also what we need to study in the future.

Finally, we adopt a light-weight model. When the input image size is $640* 480$, the model's floating point operations (FLOPs) and parameters are 2.049G and 3.241M. Inferring an image on the NVIDIA GTX 1080 GPU takes an average of 79ms, which is much lower than subsequent segmentation and recognition algorithms. Therefore, the performance loss of processing low-quality images in subsequent processes can be reduced.

\subsection{Comparison with hand-crafted factor based methods}

We also compared with single factor method which is a simplified version of multi-step method based on hand-crafted quality metrics. Sharpness, iris size, dilation, gray level spread, and usable area are selected for analysis because these factors have a great impact on recognition performance as mentioned in the IQCE challenge \cite{Tabassi_Grother_Salamon_2011}.

Sharpness $S$ uses Tenengrad focus measure operator

\begin{equation}
   S = \frac {1} {hw} \sum_x ^ w \sum_y ^ h \sqrt {G_x * I (x, y) + G_y * I (x, y)}
\end{equation}

\noindent Where $ h $ and $ w $ are the height and width of the image $ I (x, y) $, respectively, and $ G_x $ and $ G_y $ are the convolution kernels of the Sobel operator.

Iris size is defined as the number of pixels across the iris radius $ R_{iris} $, when the iris boundary is modeled by a circle.

Dilation $D$ is defined as the pupil radius $ R_{pupil} $ and the iris radius $ R_{iris}$

\begin{equation}
   D = \frac {R_ {pupil}} {R_ {iris}}
\end{equation}

Gray level spread $G$ describes the grayscale distribution of the iris area, which is defined as the information entropy of the pixel value

\begin{equation}
   G =-\sum {p_ilog{p_i}}
\end{equation}

\noindent$ p_i $ is the ratio of the number of pixels in the $ i $ th gray level to the total number of pixels. A high gray level spread image is properly exposed with a wide and well distribution of intensity values.

Usable area is defined as the percentage of iris that is not occluded by eyelash, eyelid, specular reflections and ambient specular reflections.

\begin{figure}[h]
   \begin{center}
      \includegraphics[width=0.32\linewidth]{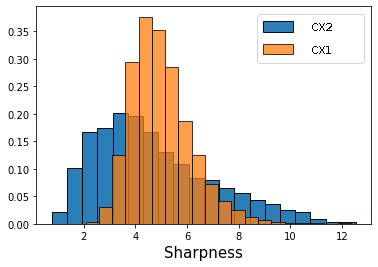}
      \includegraphics[width=0.32\linewidth]{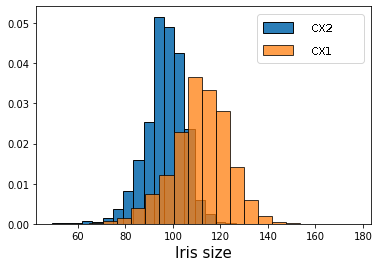}
      \includegraphics[width=0.32\linewidth]{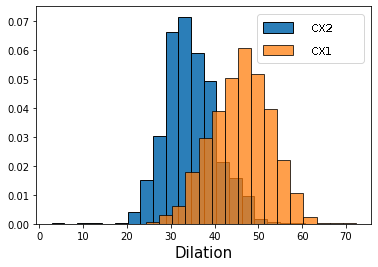}
      \includegraphics[width=0.32\linewidth]{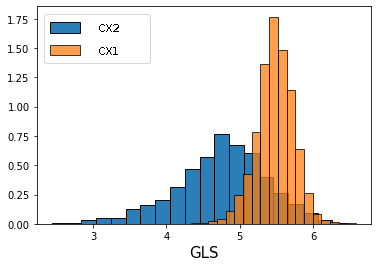}
      \includegraphics[width=0.32\linewidth]{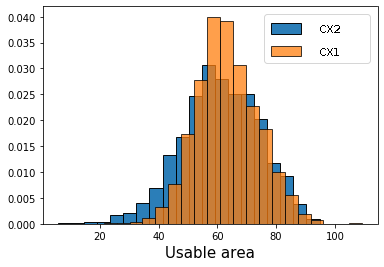}
   \end{center}
   \caption{Quality factors distribution}
   \label{fig:distribution}
\end{figure}

First, we evaluated the proposed quality metrics on two training subsets and plotted their distributions as shown in Figure \ref{fig:distribution}. And a strategy is used to determine the threshold based on the distribution. This strategy discards images with quality scores outside the range of $[\mu - \delta,\mu + \delta)$, where $\mu$ is the average of the quality metrics on the training set and $\delta$ is the threshold change.

\begin{table}
   \begin{center}
      \begin{tabular}{|l|c|c|c|}
         \hline
         Quality score     & LCC                  & SROCC                & MSE (1e-6) \\
         \hline
         Sharpness         & 0.10 / 0.28          & 0.10 / 0.27          & -          \\
         Iris size         & 0.34 / 0.11          & 0.34 / 0.21          & -          \\
         Dilation          & -0.01 / 0.14         & -0.00 / 0.17         & -          \\
         Gray level spread & -0.20 / 0.14         & -0.19 / 0.15         & -          \\
         Usable area       & 0.00 / 0.14          & -0.01 / 0.12         & -          \\
         \textbf{Ours}     & \textbf{0.78 / 0.60} & \textbf{0.78 / 0.60} & 2.2 / 8.6  \\
         \hline
      \end{tabular}
   \end{center}
   \caption{Correlation analysis between prediction / hand-crafted quality metrics and DFS}
   \label{tab:correlation_analysis}
\end{table}

\begin{table*}[h]
   \begin{center}
      \begin{tabular}{|c|c|c|c|c|c|c|c|c|c|c|}
         \hline
         EER@IRR           & \multicolumn{2}{c|}{0} & \multicolumn{2}{c|}{0.25} & \multicolumn{2}{c|}{0.5} & \multicolumn{2}{c|}{0.75} & \multicolumn{2}{c|}{0.95}                                         \\ \hline
         dataset           & CX1                    & CX2                       & CX1                      & CX2                       & CX1                       & CX2   & CX1   & CX2   & CX1   & CX2   \\ \hline
         DFS               & 12.41                  & 28.5                      & 2.91                     & 15.23                     & 0.31                      & 3.92  & 0.00  & 0.56  & 0.00  & 0.00  \\ \hline
         Ours              & 12.41                  & 28.5                      & 6.52                     & 19.89                     & 1.92                      & 14.06 & 1.44  & 7.09  & 0.40  & 5.14  \\
         Sharpness         & 12.41                  & 28.5                      & 11.99                    & 25.48                     & 11.40                     & 24.75 & 10.58 & 23.48 & 10.63 & 25.08 \\
         Iris size         & 12.41                  & 28.5                      & 10.04                    & 26.48                     & 8.67                      & 27.93 & 6.14  & 30.00 & 5.81  & 27.85 \\
         Dilation          & 12.41                  & 28.5                      & 12.42                    & 26.05                     & 12.81                     & 26.76 & 13.21 & 22.79 & 9.61  & 17.12 \\
         Gray level spread & 12.41                  & 28.5                      & 12.68                    & 28.92                     & 13.02                     & 29.27 & 14.57 & 27.10 & 18.21 & 5.94  \\
         Usable are        & 12.41                  & 28.5                      & 10.93                    & 26.54                     & 11.49                     & 26.88 & 10.75 & 27.15 & 4.68  & 32.29 \\ \hline
      \end{tabular}
   \end{center}
   \caption{Performance of the quality assessment algorithm on CASIA-Iris-Complex}
   \label{tab:irr-eer_score}
\end{table*}

After that, we also draw the IRR-EER curve of single factor method for each factors, as shown in the Figure \ref{fig:eer_irr_hd}. It can be found that these curves basically decrease significantly after the inflection point of the benchmark, which shows that relying on a single quality metric for quality assessment requires a large number of images to be discarded to ensure the performance of the iris recognition algorithm. At the same time, we noticed that the changes of the same factor curve are different on two subsets, which means that this method is not generalizable. In practice, it is difficult to design different thresholds or fusion weights in different environments.

Finally, we also compared the single-factor method with our method. It can be seen in Figure \ref{fig:eer_irr_hd} that the curve of our model is below the curves of all single-factor methods. It can be seen in Table \ref{tab:irr-eer_score} that the EER of some factors be similar to our method when the IRR is above 0.75. Further, we analyzed the relationship between the quality score of single factor method and DFS, as shown in Table \ref{tab:correlation_analysis}. Most of the quality scores have a low correlation with DFS, which also shows that the hand-crafted quality metric is not designed for better recognition.

\begin{figure}[h]
   \begin{center}
      \subfigure[]{
         \includegraphics[width=0.8\linewidth]{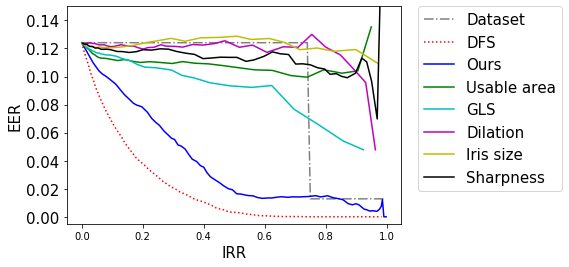}
         \label{fig:eer_irr_hd}
      }
      \subfigure[]{
         \includegraphics[width=0.8\linewidth]{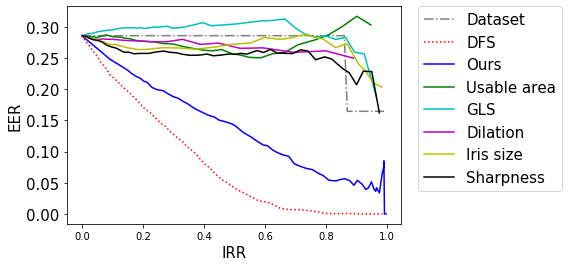}
         \label{fig:eer_irr_hd}
      }
   \end{center}
   \caption{EER-IRR curve of hand-crafted factor based method, (a) is CASIA-Iris-CX1, (b) is CASIA-Iris-CX2.}
   \label{fig:eer_irr_hd}
\end{figure}

We did not conduct further experiments because both multi-step method and fusion method are based on a single-factor architecture. The multi-step method is actually a cascade of single factors, so the IRR will not be smaller than the single factor method while ensuring the same recognition performance. The fusion method is to fuse single factors through a pre-designed function and calculate a final quality score. If the prediction ability of a single factor is weak, the prediction ability of the fusion method will not be significantly improved.

According to the experiments, we found that using DFS as the iris image quality metric can improve the performance of the iris recognition system while reducing the number of discarded images. And our proposed attention-based quality assessment model can effectively predict the DFS value and ensure the running speed, which is more effective than the traditional hand-crafted factor based assessment methods.

\section{Conclusions}

In this paper, we propose a recognition-oriented iris image quality assessment method that utilizes the distance of iris embeddings in feature space as image quality metric. In order to predict the distance in feature space, a neural network based on attention mechanism is introduced. Then, a self-established database containing both ideal and non-ideal iris images is released in this paper to study the problem of iris image quality assessment and robust iris localization, segmentation and recognition algorithms. We use this database to demonstrate that our proposed quality assessment method based on DFS can effectively improve iris recognition performance while discarding as few image as possible during preprocessing. It is noted that our attention based neural network predict DFS in real-time and more effective than hand-crafted factor based quality assessment methods.

Furthermore, most biometric recognition algorithms map biometric samples to a feature space and perform further recognition, verification, or clustering. Therefore, the DFS-based biometric sample quality assessment method proposed in this paper can also be applied to those fields. In addition, the proposed quality metric can also be used as the supervision for image enhancement, reducing image rejection rate and improving the performance of image understanding algorithm through visual information enhancement. Compared with traditional hand-crafted factors methods which regarded quality assessment and recognition as separate tasks, our proposed method is a trial to bridge the gap between the quality assessment and recognition.

~\\\noindent\textbf{Acknowledgments} This work is supported by National Key Research and Development Project (Grant No.2017YFB0801900), National Natural Science Foundation of China (Grant No. U1836217, 61427811), Tianjin Key Research and Development Project (Grant No.17YFCZZC00200)  .

{\small
\bibliographystyle{ieee}
\bibliography{submission_example}
}

\end{document}